\documentclass[prl,aps,twocolumn,showpacs,floatfix]{revtex4}

\usepackage{amssymb}
\usepackage{amsmath}
\usepackage{mathtools}
\usepackage{amsfonts}
\usepackage{graphics}
\usepackage{color}
\usepackage{verbatim}
\usepackage[normalem]{ulem}	

\begin{document}
\title{High tunability of the transport properties in macroscopically in-plane modulated two-dimensional system}

\author{ A.V. Shupletsov$^{a}$\thanks {e-mail:
husderbec@mail.ru}, A.Yu. Kuntsevich$^{a,b}$,
M.S. Nunuparov$^{c}$, A.L. Rakhmanov$^{a,d,e,f}$  K.E. Prikhodko$^{g}$
}

\affiliation{$^a$P.N. Lebedev Physical Institute of the Russian Academy of Science, 53 Leninskiy prospekt, Moscow, 119991, Russia}
\affiliation{$^b$National Research University Higher School of Economics, Moscow, 101000, Russia}
\affiliation{$^c$Prokhorov General Physics Institute of the Russian Academy of Sciences, Moscow, 119991, Russia}
\affiliation{$^d$Dukhov Research Institute of Automatics, Moscow, 127055, Russia}
\affiliation{$^e$Institute for Theoretical and Applied Electrodynamics, Russian Academy of Sciences, Moscow, 125412, Russia}
\affiliation{$^f$Moscow Institute of Physics and Technology, Dolgoprudny, Moscow Region, 141700, Russia}
\affiliation{$^g$National Research Center Kurchatov Institute, 1 Akademika Kurchatova pl., Moscow, 123182, Russia}

\begin{abstract}
Gate-controllable two dimensional systems with in-plane modulation of properties could serve as highly tunable effective media. Intuitively, such systems may bring novel functionality provided that the period of the lateral modulation is much less than the relevant scattering lengths (mean free path, coherence length etc.). Our work experimentally demonstrates the opposite, disordered limit of such system, defined in the macroscopically modulated metal-oxide-semiconductor structure. The system consists of parent two-dimensional gas with periodic array of islands (dots/antidots), filled with two-dimensional gas of different density, and surrounded by depletion regions (shells). Carrier densities of both parent gas and islands are controlled by two independent gate electrodes, allowing us to explore a rich phase diagram of low-temperature transport properties of this modulated two-dimensional system, resembling various transport regimes: insulating, shell-dominated, gas-dominated, island-dominated. These regimes can be identified by various Hall resistance and its magnetic field dependence, temperature dependencies of the resistivity, and Shubnikov-de Haas patterns. We also suggest the theoretical approach for description of such inhomogeneous but periodical systems. Theory based on the classical mean field approach qualitatively describes our system as theoretical dependencies reproduce the main features of the experimental behavior of the effective Hall concentration from gate voltage. Thus, our work demonstrates feasibility of the macroscopically inhomogeneous two-dimensional system as a tunable platform for novel physics and proposes the approach for the theoretical description of such systems.
\end{abstract}

\pacs{73.50.Jt ,73.40.Qv, 85.30.Tv}

\maketitle

\section{Introduction}
Two-dimensional electron systems (2DES) are convenient platforms for numerous physical experiments and applications. Adding up lateral modulation turns a system into two-dimensional metamaterial and opens additional functionality, like gate-tunable superconductivity in a lattice of superconducting tin islands on graphene \cite{ControlSC}, discovery of correlated state and superconductivity in magic-angle twisted bilayer graphene \cite{matblg1,matblg2}, experimental observation of Holfstatder's butterfly \cite{Hofstadter}, commensurability effects in semiconducting quantum wells with lateral modulation \cite{Commensurability}, selectivity to circularly polarized light in the chiral laterally modulated structures \cite{chiral} etc. In all these examples the modulation period is smaller than the relevant length, e.g. mean free path or coherence length, otherwise the effects of periodical modulation would be damped.

On the other hand, even if the latter condition is violated,  the modulated system remains to be a regular effective medium anyway. We address a question to what extent should one expect the emergence of new phenomena there? Conductance of such effective medium is very sensitive to electronic properties and could serve therefore as a convenient indicator.
In this paper we examine the conductive properties in the disordered and yet not insulating limit of macroscopically modulated gate-tunable array of islands (dots/antidots) within 2D electronic system, realized in the archetypal Si-MOSFET platform.
This system is somewhat similar to granular materials, studied broadly in the past \cite{Abeles} both theoretically and experimentally. The studied array of islands differs from granular systems by:
(i) complete two-dimensionality and tunability of both parent electron gas and islands;
(ii) periodicity, i.e. absence of randomness in positions of islands;
(iii) smooth transition regions (larger than mean free path) between parent gas and islands.

So far transport studies of lithographically modulated semiconducting two-dimensional systems were focused either on clean systems (where mean free path is larger than the period of modulation and all studied phenomena are essentially ballistic \cite{Weiss, weiss2,kozlov1,kozlov2,tsukagoshi}) or to Aharonov-Bohm/Altshuler-Aharonov-Spivak oscillations \cite{AharonovBohm,AlshulerAharonovSpivak}, i.e. coherent low-temperature mesoscopic effects \cite{nihey, iye, yagi}. All these phenomena are essentially nano-scale. We should also mention a group of a papers \cite{dorn,minkov,staley,goswami, tkachenko}, where percolation and transition to localization phenomena in the arrays of dots/antidots were explored.

Arrays of {\it macroscopic} (i.e. micrometer size) islands should address essentially the classical physics. Macroscopic means that the mean free path($<50$ nm in our case) and coherence length ($\sim$ 300 nm at 2K) are smaller than the period of the structure and size of the islands (in our case 5 and 2.5 $\mu$m respectively). To the best of our knowledge, magnetotransport properties of such system (array of depleted antidots) so far were reported only by us in Ref.\cite{kuntsevich}, where Hall resistance was shown to be nonlinear function of magnetic field due to current redistribution in magnetic field. Present study qualitatively extends those first measurements, by adding a new parameter, i.e. electron density in the islands. From the transport studies we explore island density/2DES density phase diagram of this effective media. We reveal and explain qualitatively 2DES-dominated, island-dominated and shell dominated phases, highlight the role of inhomogeneities in 2D-metal-to-insulator transition. Our data indicate weak-localization related reason for low-field Hall nonlinearity and novel effect in the Shubnikov-de Haas oscillation regime: Zeeman splitting of the resistivity minima. Also in this paper we find the analytical expression for the effective conductivity of the model system using classical mean field approach. Considered system differs from the experimental one primarily by the neglect of the transitional regions between islands and parent gas and quantum corrections in conductivity. The theoretical model qualitatively describes the experimental system reproducing the non-linear dependence of Hall conductivity from the voltage on the gate above parent gas.

\section{Samples used}
\begin{figure}
\centerline{\includegraphics[width=\linewidth]{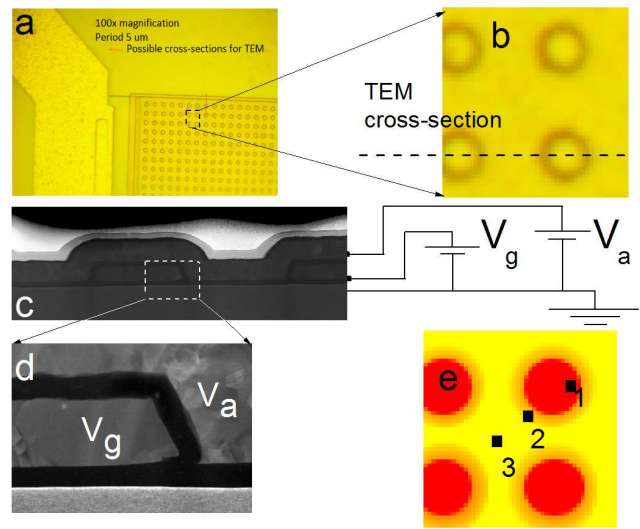}}
\begin{minipage}{3.2in}
\caption{(Color online) (a) Optic image of the corner of the AA (100x magnification), (b) zoom-in of image (a) with direction of slice for TEM, (c) TEM image with scheme of the gating, (d) TEM image of the border of the island (on the right), S2DG (on the left) and shell (between), (e) image (b) with signed areas (1-island, 2-shell, 3-SD2G).}
\label{Samples}
\end{minipage}
\end{figure}

We used Si-MOSFETs structures with lithographically defined antidot array (AA) (for simplicity, we call islands antidots though they can be dots), with TEM cross-section and gate connection shown schematically in fig.\ref{Samples}. The transport current flows in the inversion layer at the interface between Si substrate and oxide. Voltages applied to two electrically decoupled gate electrodes independently control the density of the electrons (i) inside the antidots ($V_{a}$) and (ii) in the surrounding 2D gas (S2DG) ($V_{g}$).  Panels \textbf{a}, \textbf{b} show optical images of the sample. Diameter of the antidots is 2.5 $\mu m$, lateral period $d$ of the structure is 5 $\mu m$ so that transport between them is diffusive ($l \ll d$ where $l \sim 50$ nm is mean free path in the highest mobility samples) and possible coherent effects are negligible ($l_\phi < d$ where $l_\phi < 500$nm is coherence length in studied temperature range). The AA has a Hall-bar shape with lateral dimensions 0.4 mm x 0.4 mm. 

The cross-section thin lamella for TEM studies was cut out from the surface region (shown by dashed line in panel \textbf{b} of Fig.\ref{Samples}) of the sample  using FEI Helios NanoLab 650 focused ion beam. The STEM images (see example in Figs.\ref{Samples}\textbf{c} and \ref{Samples}\textbf{d}) were obtained using FEI Titan 80-300 microscope at the electron energy of 200 keV.

The structure of our sample is following: bottom layer in gray color - single crystalline (001) Si substrate; the dark color corresponds to $SiO_2$, trapezoidal-shaped polycrystalline heavily doped Si is the gate of the antidots, the rest polycrystalline heavily doped Si (gray color above $SiO_2$) is the S2DG gate. 
Panel \textbf{d} shows the zoom in of the edge of the antidot. It is seen that the oxide layer becomes thicker closer to the edge of the S2DG. This leads to lower density of electrons in the  domains underneath. Moreover, gate electrodes are separated by oxide so that between antidots and S2DG there is an area where density is expected to be low. We call these transition regions \textit{shells}.
The panel \textbf{e} shows the same spot as panel \textbf{b} with all mentioned above areas in color.

Multiple chips of the same design were fabricated on the same wafer. Probably due to inevitable temperature gradients during the fabrication, AA on different chips demonstrated different low-temperature transport properties. In particular, peak mobility varied by an order of magnitude (see Results section). 

\section{Results}
Magnetoresistance measurements were performed in the temperature range 0.3-8 K using Cryogenics 21T/0.3 K and CFMS 16T/1.8K systems. AC transport current was fixed at value 100 nA to avoid overheating. All measurements were carried out in the frequency range 13-18 Hz using a standard 4-terminal technique with a lock-in amplifier. In order to compensate for contact asymmetry, magnetic field was swept from positive to negative values and with resistance per square (Hall resistivity) data being then (anti)symmetrized.

The properties of Silicon-based 2D systems are known to be strongly dependent on the mobility of carriers. In high-mobility uniform systems ($\mu\gtrsim 10000$cm$^2/$Vs) metallic behavior of resistivity and metal-insulator transition can be realized \cite{MIT}. In contrast, low-mobility Si-MOSFETs do not demonstrate a stark metallic temperature dependence of the resistivity. Also, for high-mobility samples Shubnikov-de Haas oscillation(SdHO) patterns allowed to resolve the carrier density value $n_{SdH}$.

The experiments were carried out on several samples with effective peak mobility of electrons in AA in wide range from 400 to 5000 cm$^2/$Vs. Despite this spread of mobilities, most of the observed phenomena were shown up in all samples. The mobility had impact only on the magnitude of the corresponding effects. We demonstrate the data from the representative high mobility sample AA1 and low mobility sample AA2.

All measurements are made in the regime of highly conductive media. Indeed, measured resistance per square (that is always elevated with respect to the S2DEG local resistivity due to bottleneck effect) is lower than the resistance quantum $h/e^2\sim 25.8$ kOhm. Therefore the quasiclassical treatment of the transport is applicable.

\subsection{Effective density}
We straightforwardly characterize this effective medium by effective Hall density ($n_{eff}\equiv{[eR_{xy}(B=1{\rm T})]^{-1}}$) and effective carrier mobility ($\mu_{eff}\equiv{(n_{eff}e\varrho)^{-1}}$). Here and further $\varrho$ is the measured resistance per square. The effective density and mobility were calculated from the resistance per square and Hall resistivity at 1 and -1 T.

\begin{figure}[ht]
\centerline{\includegraphics[width=\linewidth]{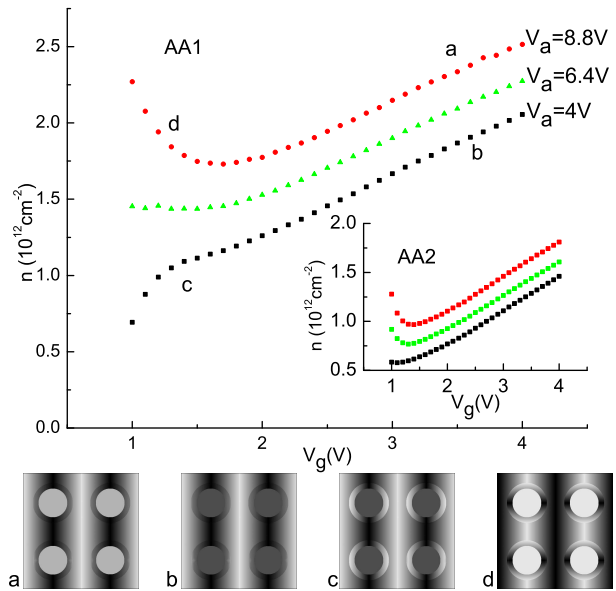}}
\caption{(Color online) The Hall density at T=1.8K for sample AA2 vs S2DG gate voltage for three representative island gate voltages. Inset shows the similar data for the high-mobility sample AA1. In panels (a)-(d) the higher electron density the lighter area. Panels (a)-(d) correspond to the domains of the voltages designated by the same letter on the graph.}
\label{n(Vg)}
\end{figure}

We analyzed the $n_{eff}$ dependency on $V_g$ and $V_a$.
In uniform Si inversion layers electron density is roughly proportional to $(V_g-V_{th})$ \cite{Ando}, where $V_{th}$ is a threshold voltage, which is usually small and originates from charge stored in oxide and the difference of work functions of the gate and 2D system. Experimentally observed $n_{eff}(V_g)$ dependencies (for three various $V_a$ values, shown in fig.\ref{n(Vg)}) are in contrast with this expectation. The reason for the deviations is artificial non-uniformity of the system. Such behavior reflects different regimes of transport current flow distribution. We distinguish the ranges of gate voltages that correspond to various current density distribution (schematically shown  by letters (a)-(d) in the main panel and also in the corresponding panels  under the graph in fig.\ref{n(Vg)}). The higher transport current density is shown by lighter color.

For $V_g$ high enough (figures \ref{n(Vg)}(a) and \ref{n(Vg)}(b)), S2DG is very conductive because of high electron density. Due to edge effects and larger gate-to-2DEG distance, shells have lower electron density and hence smaller conductance. Therefore, transport current flows predominantly through the S2DG and Hall effect, i.e. $n_{eff}(V_g)$, is determined by its density. It means that the islands have small impact on $n_{eff}(V_g)$ dependence.

For small values of $V_g$ (figures \ref{n(Vg)}(c) and \ref{n(Vg)}(d)) the S2DG density and conductivity decreases and contribution of islands to the transport rises. Increasing the $V_a$ value makes the islands much more conductive than S2DG. Therefore transport current prefers to flow through islands and minimizes the path through the S2DG. Thus the effective density $n_{eff}$ increases (relatively to density defined by $V_g$) since the Hall voltage is determined by the islands. As $V_g$ increases, the contribution of depleted S2DG rises leading to the drop of the $n_{eff}$ (figure \ref{n(Vg)}(d)).

For $V_g$ and $V_a$ low enough, both S2DG (unlike case \textbf{b}) and islands (unlike case \textbf{d}) are poorly conductive. Low conductance of both regions force transport current to flow through the whole perimeter of the shell. This leads to the elevated role of the low-density shells and the visible increase of the Hall voltage, i.e. drop of $n_{eff}$ value (case \textbf{c} on the fig.\ref{n(Vg)}).

The effective density data, shown in fig.\ref{n(Vg)}, demonstrating an enhanced drop in low-$V_a/$ low-$V_g$ region, were obtained for low-mobility sample AA2. For high-mobility sample AA1 (inset to fig.\ref{n(Vg)}), despite the absence of the drop, a similar tendency is clearly seen: $n_{eff}$ value decreases with $V_g$ growth at high $V_a$ and this effect vanishes as $V_a$ is lowered. This data show that the effective density in the macroscopically inhomogenious systems follows the same physics irrespectively of mobility.

\subsection{Magnetoresistance and Hall measurements}

Thus, we established different regimes of current transport in artificially inhomogeneous tunable media. In order to explore the differences between the regimes \textbf{b}, \textbf{c}, and \textbf{d} (here and further designations are taken from fig.\ref{n(Vg)}) we performed more detailed magnetotransport measurements. We chose Hall coefficient ($R_{xy}/B$) to visualize the difference between AA and uniform system, where $R_{xy}/B$ is roughly field-independent.

\begin{figure*}
\begin{minipage}[h]{0.49\linewidth}
\centerline{\includegraphics[width=\linewidth]{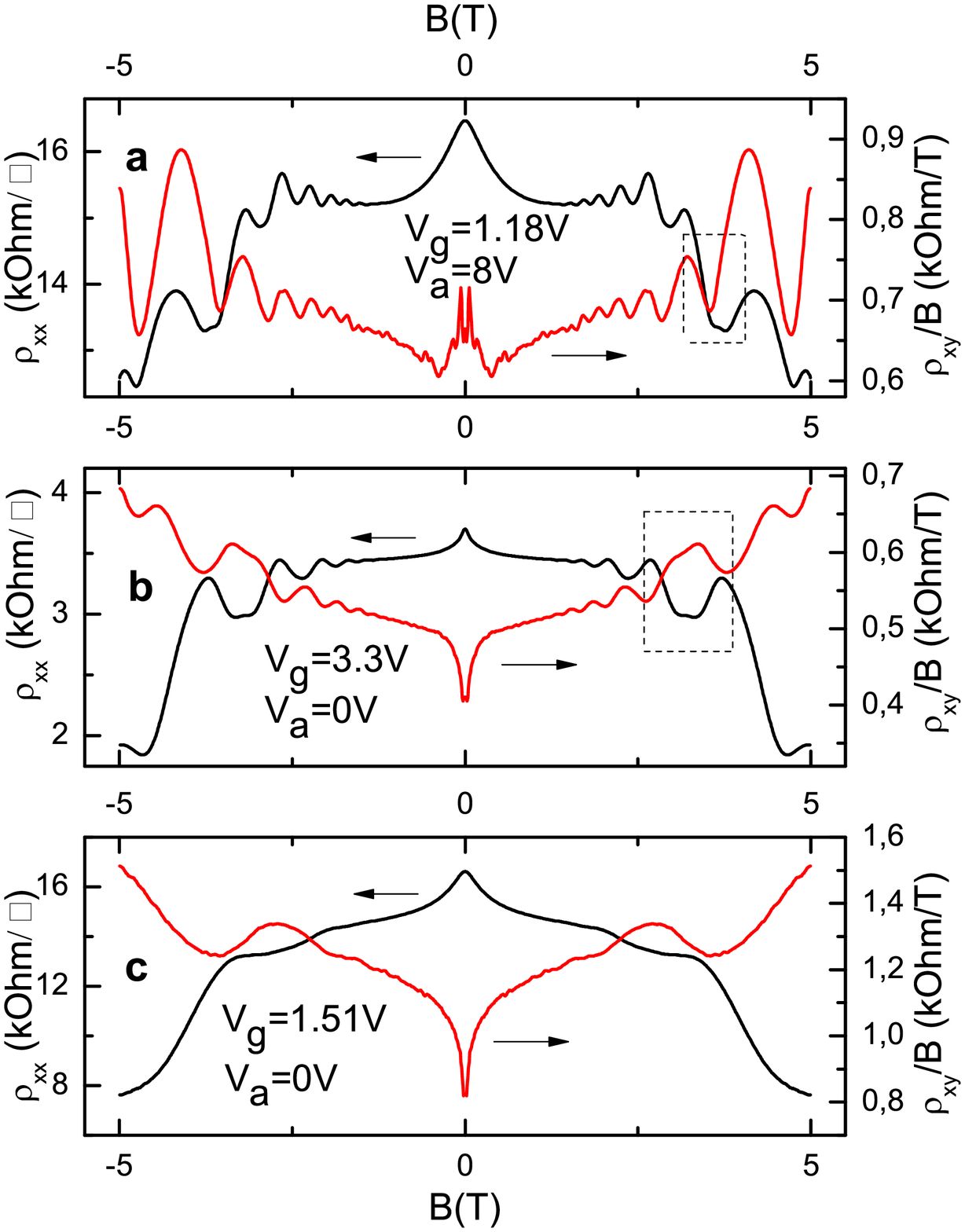}}
\end{minipage}
\hfill
\begin{minipage}[h]{0.49\linewidth}
\center{\includegraphics[width=\linewidth]{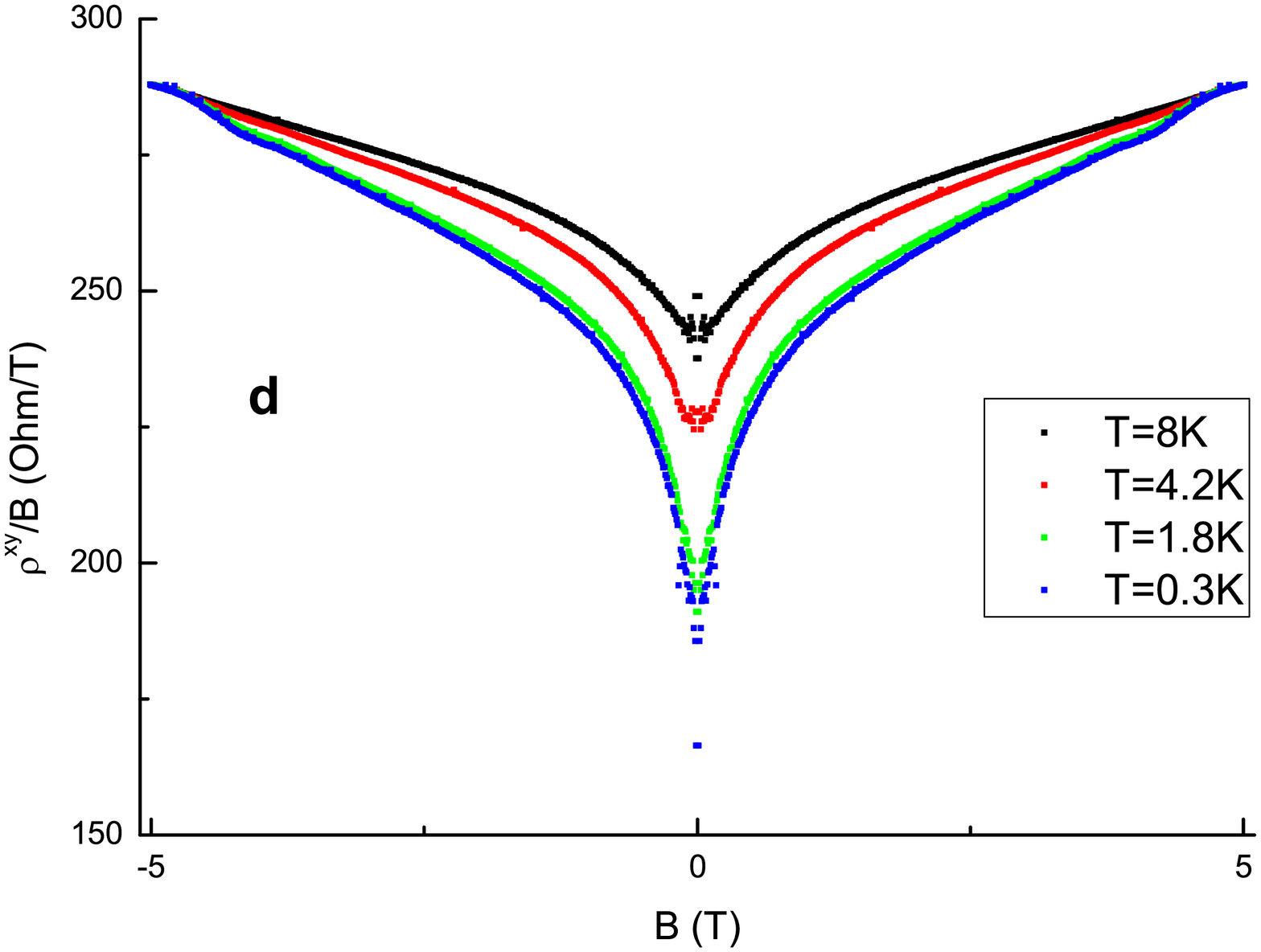} \\
\includegraphics[width=\linewidth]{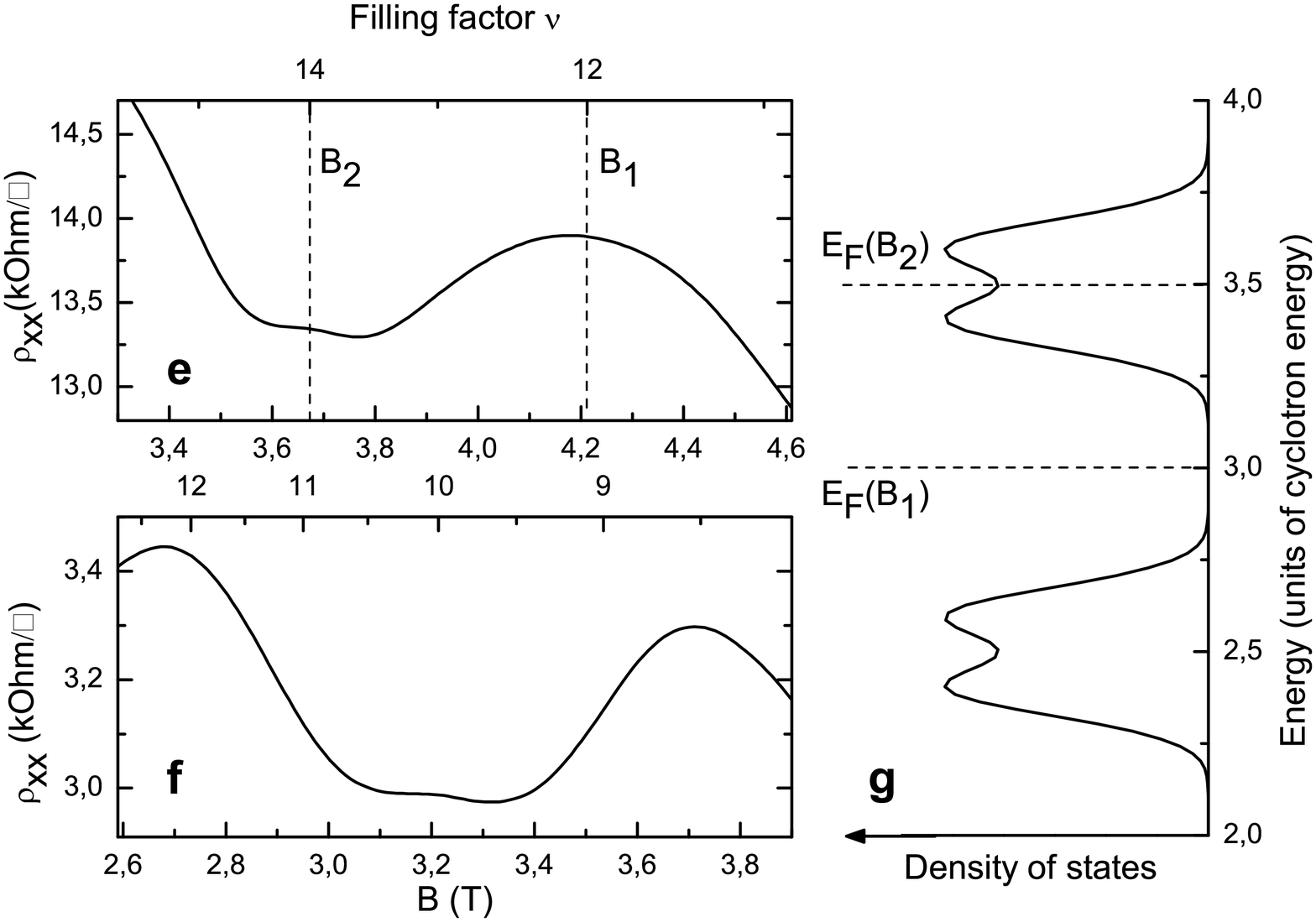}}
\end{minipage}
\caption{(Color online) Magnetoresistance (black curves) and Hall coefficient (red curves) of sample AA1 at T=0.3K in regime of current through antidots (a), current through S2DG (b) and elevated role of shells (c). (d) Hall coefficient of sample AA2 vs magnetic field for four different temperatures. For convenience, curves shifted such that their edges (at B=5T) coincide (curve for 0.3K remained unchanged). (e) and (f) are enlarged areas from panels (a) and (b), respectively, shown by dashed rectangles which demonstrate the splitting of minima of magnetoresistance. (g) Schematics of Zeeman-splitted Landau levels in density of states vs energy diagram.  Fermi levels for magnetic fields $B_1$ and $B_2$ (indicated in panel (e)) are shown by dashed lines. }
\label{R(B)}
\end{figure*}

Fig.\ref{R(B)}(a) shows magnetoresistance ($\varrho$) and Hall coefficient in the regime \textbf{d} where transport is dominated by islands. Though the effective $\varrho$ is about 15 kOhms (i.e. $\sim e^2/h$), pronounced Shubnikov-de Haas oscillations (SdHO) are observed due to the high mobility electron gas in the islands. Electron density obtained from SdHO ($n_{SdH}\approx 1.4\times10^{12}$cm$^{-2}$) is higher than the Hall density ($n_{eff}\approx 1\times10^{12}$cm$^{-2}$) because the latter is affected also by S2DG bottlenecks. Hall coefficient is a non-monotonic function of magnetic field with a \textit{maximum} at $B=0$.

For comparison in fig.\ref{R(B)}(b) we show magnetoresistance and Hall coefficient of the system in regime \textbf{b} with $V_a=0$. $V_g$ value was adjusted to make $n_{eff}$ approximately equal to the value from fig.\ref{R(B)}(a). Effective $\varrho\approx3$kOhms value is about 5 times less because S2DG in this case is well-conductive and transport current bypasses the depleted regions. SdHO are also observed with $n_{SdH}\approx 0.9\times10^{12}$cm$^{-2}$ comparable to $n_{eff} \approx 1.2\times10^{12}$cm$^{-2}$. At $B=0$ Hall coefficient in this case has {\it minimum}.

Finally, fig.\ref{R(B)}(c) shows magnetoresistance and Hall coefficient in low-density regime somewhere between \textbf{b} and \textbf{c}. The gate voltages were adjusted to make $\varrho$ approximately equal to the value from fig.\ref{R(B)}(a). The behavior of the transport is completely different from fig.\ref{R(B)}(a) and qualitatively similar to fig.\ref{R(B)}(b) without SdHO. Hall coefficient has {\it minimum} at zero field. This data straightforwardly demonstrates that contrary to non-modulated 2DES, the magnetotransport reflects complexity of carrier density redistribution and is not determined by the value of the effective resistivity.

The common tendency for all Hall coefficient data is the growth with the magnetic field. In homogeneous system Hall coefficient is constant and directly corresponds to electron density $R_{xy}/B=1/ne$. In the studied system there are regions with different densities. For Si MOSFETs it is known that electron mobility is density-dependent ($\mu$ generally  grows with $n$, then reaches a steep maximum and decreases slowly for very large carrier densities) \cite{Ando}. In magnetic field the longitudinal conductivity $\sigma_{xx}$ decreases $\propto (1+(\mu B)^2)^{-1}$, i.e. the higher the mobility, the faster the decrease. Thus, with increasing the field the conductivity of low-density regions decreases slower than the one of high-density regions. Since the current prefers to flow through high-conductive regions, with increasing field current redistributes so that the role of low-density low-mobility regions increases. Therefore, Hall coefficient should rise, in agreement with experimental data. Exactly this mechanism was suggested in our first paper\cite{kuntsevich}.

In small magnetic field Hall coefficient experiences an abrupt feature. The bare 2D gas in Si-MOSFETs also has a small low-field Hall nonlinearity, discussed in detail in Ref.\cite{kuntsevich2013} and reported for the similar samples in Ref.\cite{kuntsevich}. However, the huge amplitude of the low-field Hall coefficient variation in Fig.\ref{R(B)} clearly identifies it with the sample nonuniformity. This huge non-linearity is one of the main observations of our paper.
Interestingly, low-field quenching of transverse magnetoresistance (and even change of its sign) has already been explored in various artificially inhomogeneous and mesoscopic systems. First experiments in 1D wires by Roukes \cite{Roukes} were further theoretically explained \cite{Beenakker} by scrambling of electron trajectories on crossroad in a place of contacts. The authors speculated that quenching is unambiguous manifestation of 1D transport. We note that all available theories in 1D or 2D systems are essentially ballistic. In further experiments with ballistic antidot arrays\cite{Weiss} the quenching of the Hall effect was also observed, although the qualitative pinball picture didn't account for attenuation of Hall coefficient. In the more recent experiments on 2D systems with AA \cite{Cross-shapedAA} the observed quenching of Hall effect was confirmed by numerical simulations, but no physical mechanism was suggested.

Our system is essentially different, because the transport is diffusive and the inhomogeneities are tunable from dots (areas of low potential, $V_a>V_g$) to antidots (areas of high potential, $V_a<V_g$). Zero-field Hall coefficient in our experiments can either grow or fall with $B$ depending on $V_g$ and $V_a$. Origin of different behavior is unclear and requires further theoretical investigation. Suppression of the zero-field Hall coefficient quenching with temperature (fig.\ref{R(B)}(d) for sample AA2) is the indicator, that this feature is related to weak-localization phenomenon.
We believe that low-field feature in Hall coefficient comes from redistribution of transport current in the regime of weak localization.
This assumption is totally nontrivial: firstly, it is a textbook knowledge that in homogenous medium weak localization does not influence the Hall resistivity\cite{AltshulerAronov} and, secondly, the relative value of the observed nonlinearity is rather high (few 10\%), larger than weak localization correction to resistivity in the bare 2D gas. Our results thus call for theoretical modeling of the weak localization in the presence of macroscopic modulation. Moreover, it might be that sample inhomogeneity is a clue to understanding the often observed and not always explained low-field feature in the other 2D systems \cite{kuntsevich2013, ovadiahu, MinkovHall}.

Another unusual, yet high-field magnetotransport effect is the splitting of the minima of the longitudinal magnetoresistance $\varrho$ (enlarged domains from figs.\ref{R(B)}(a-b) are shown in Fig.\ref{R(B)}(e-f)). As a rule, as magnetic field increases, and Zeeman term exceeds the temperature and Landau level broadening (see fig.\ref{R(B)}g for the schematics of the density of states), the resistivity maxima are split. Indeed, in uniform 2D systems in SdH domain Hall resistivity $\rho_{xy}$ is higher than $\rho_{xx}$ and the maxima of the conductivity $\sigma_{xx} = \rho_{xx}/(\rho_{xx}^2+\rho_{xy}^2)\approx\rho_{xx}/\rho_{xy}^2$ at the half-integer filling factors correspond to the maxima of the resistivity and maxima of the density of states.

In our samples, effective resistance per square $\varrho$ is higher than $R_{xy}$ in SdH domain. If the areas of antidots were just infinite barriers for electrons, it would only change the geometrical factor $w/l$ and do not turn minima to maxima. In other words, the resistance per square should increase but the $\varrho(B)/\varrho(B=0)$ ratio should remain unchanged. Meanwhile in our system $\varrho_{xx}$ {\it minima} appear to be splitted. It is worth to note that splitting is observed both in regime {\bf b} (current through S2DG) either in regime {\bf d} (current mainly through islands).

\begin{figure*}[t]
\begin{minipage}[h]{0.49\linewidth}
\centerline{\includegraphics[width=\linewidth]{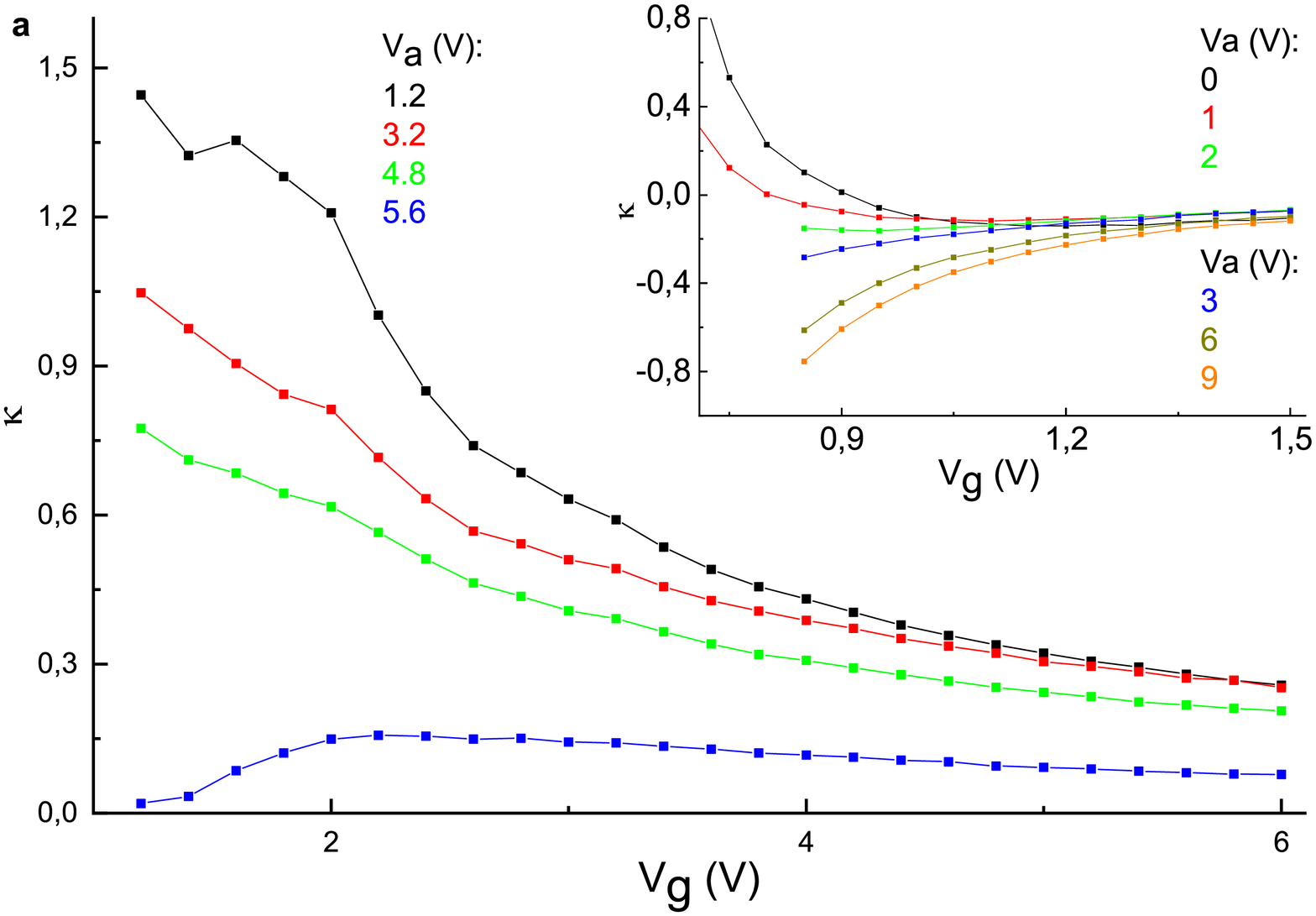}}
\end{minipage}
\hfill
\begin{minipage}[h]{0.49\linewidth}
\centerline{\includegraphics[width=\linewidth]{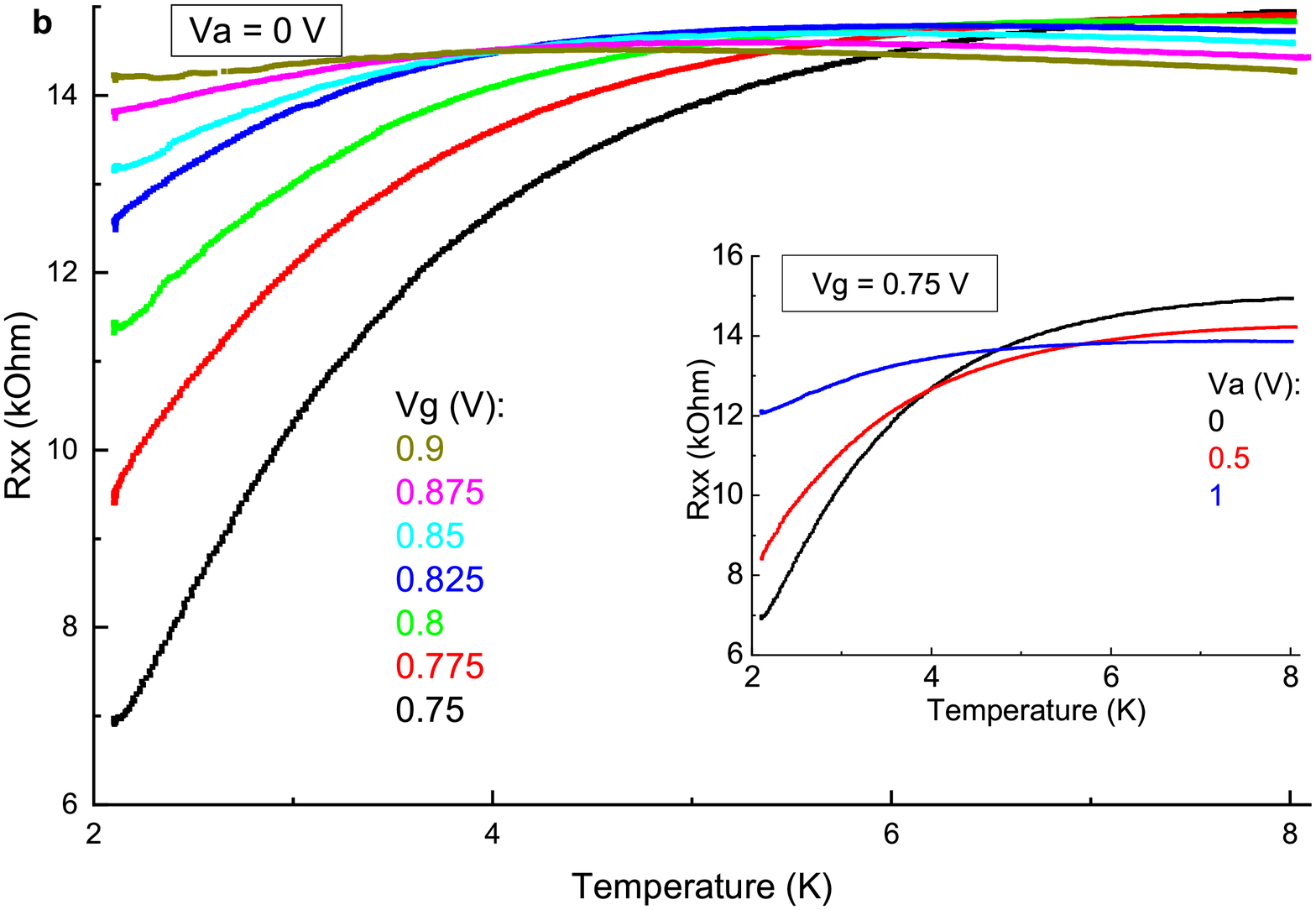}}
\end{minipage}
\caption{(Color online) (a) Relative change of resistivity of AA1 with temperature (from 1.8K to 7.4K) $\kappa$ vs S2DG voltage for different voltage on antidots gate. The same data (but for temperatures 2.1K and 8K) for low-mobility sample AA2 is shown on inset. (b) Temperature dependence of resistivity of the sample AA2 at fixed $V_a$=0V for different $V_g$. The same data at fixed $V_g$ for different $V_a$ is shown on inset.}
\label{Metallicity}
\end{figure*}

We suggest that this splitting might be explained if the equation $\sigma^{eff}_{xx} = \varrho/(\varrho^2+R_{xy}^2)$ holds correct for the resistance per square. Then $\sigma^{eff}_{xx}\approx 1/\varrho$ and conductivity maxima (coinciding with the maxima of Zeeman-splitted density of states at Fermi level, shown in Fig.\ref{R(B)}g) correspond to the resistance per square minima. This suggestion is not expected to be valid because conductivity and resistivity are local properties, whereas the resistance per square is the macroscopic characteristic of the sample. In other words, effective conductivity approach is surprisingly applicable not locally but rather to the overall system.

\subsection{Metallic behavior of resistivity}

High mobility Si-based 2D systems are also remarkable by ``metallic'' resistivity behavior ($d\rho/dT >$ 0) and metal-insulator transition \cite{MIT}. These phenomena were intensively investigated during last two decades. They are shown to occur due to interplay of strong electron-electron interactions and localization, however the exact mechanism is yet debated \cite{spivakVK, dobrosavlevic, finkelsteinRG, french, meir, pudalovNonUniform, gold}.

Since in some of these models the system was believed to be essentially non-uniform at the microscale  \cite{spivakVK, pudalovNonUniform, meir} we decided to examine how the artificially tunable inhomogeneity in our system will affect 2D ``metallicity''.

We should note that even in non-modulated Si-based 2DES a valuable metallicity (2-5 times growth of the resistivity from $\sim 1$K to $\sim 10$ K ) emerges only if the peak mobility is rather large ($\mu>1$ m$^2$/Vs). In this case the strength of metallicity grows as density decreases and eventually quenches at metal-insulator transition point.
If the peak mobility is low, than low densities are not achieved and magnitude of resistivity variation with temperature becomes small or even slightly negative.

In order to quantify ``metallicity'' experimentally we took the relative variation $\kappa\equiv (\varrho_{7.4}-\varrho_{1.8})/ \varrho_{1.8}$, where $\varrho_{1.8}$ and $\varrho_{7.4}$ values were measured at experimentally convenient temperatures 1.8K and 7.4K respectively.
Thus, $\kappa$ never drop below -1, and relatively big positive values of $\kappa$ correspond to strong ``metallic'' behavior and negative values - to insulator. Figure\ref{Metallicity}\textbf{(a)} shows $\kappa$ versus $V_g$ dependence for different values of $V_a$ for high-mobility sample AA1. The inset shows a similar series of $\kappa(V_g)$ dependencies for low-mobility sample AA2. Insignificant distinction is that temperature reference points used for sample AA2 were 2.1K and 8K, respectively. This difference is connected only with experimental conveniences.

At high values of $V_g$, when the system is deep in the conductive domain $\kappa$ tends to zero for both low and high-mobility samples.
This behavior is caused by (i) weakening of  electron-electron interactions at elevated densities and (ii) domination of the S2DG in conductance of the system, i.e. transport properties of antidot array for large $V_g$ are equivalent to bare 2D gas, as expected.

For small values of $V_g$ $\kappa$ depends dramatically on the value of $V_a$ and on the mobility of the sample.

For high-mobility sample and small values of $V_a$ there is a strong ``metallic'' conductivity: $\kappa$ is positive, quite large (about 1.5), and drops monotonically with $V_g$, as it should be for bare 2D gas in Si-MOSFET\cite{MIT}, because the islands areas are out of the game. However for high values of $V_a$ ``metallic'' conductivity becomes suppressed for all values of $V_g$ and $\kappa$ for sample AA1 becomes non-monotonic and goes to zero at small $V_g$.

For low mobility sample AA2 as $V_a$ increases weakly positive $\kappa$ for low $V_g$ turns to negative. Fig.\ref{Metallicity}(b) represents the temperature dependence of the resistivity of low-mobility sample. On the main graph dependencies for different S2DG voltage $V_g$ and same antidots voltage $V_a$ are shown to demonstrate the degeneracy of metallic behavior with increase of $V_g$. The same tendency with $V_a$ increase at fixed $V_g$ is reflected in inset. In other words filling the islands with electrons turns the system to ``insulating'' behavior, no matter how large the mobility is.

We suggest the following explanation of this phenomena. For small values of $V_a$ islands areas are ``closed'' for electrons. However for high values of $V_a$ current flows to the islands and, as result, inevitably flows through the shells. The latter have strong insulating behavior that cause the suppression of $\kappa$. Thus, we demonstrate and explain qualitatively that our effective media allows to tune 2D ``metallicity''.

This observation might also help to understand the answer to the question why the strength of the metallicity in Si-MOSFETs is the highest among other system despite the relatively low mobility. Indeed, in the highest mobility Si-MOSFETS ($\mu_{peak}\sim 3-4\cdot 10^4$cm$^2/$Vs), the resistance increases almost by an order of magnitude with temperature\cite{MIT}, and strong metallicity is observed at relatively high carrier densities (few $10^{11}$ cm$^{-2}$). In the other material systems with mobilities exceeding $10^5$ cm$^2/$Vs and much lower carrier densities ($\sim 10^{10}$ cm$^{-2}$, e.g. Si/SiGe quantum wells\cite{melnikov}, n-GaAs \cite{lilly}, p-GaAsTO\cite{proskuryakov}, etc.) the growth of the resistivity with temperature is typically smaller than a factor of 2.
All these high mobility systems have smooth impurity potential, similarly to tunable part of potential due to artificial modulation in the antidot array. This potential might be one possible mechanism for the metallicity suppression. Indeed, in low carrier density materials the relative fluctuations of charge distribution are much larger and in their role should be re-examined.

\section{Theory}
Interestingly, there is a possibility to obtain \textit{analytical} results for the conductivity of the regular array of equivalent elliptic islands embedded into a conductive matrix\cite{MFT}. For theoretical description here we consider infinite 2D array of round islands with radius $R$ and period $a$ and the main matrix (S2DG). The input parameters are magnetic field directed perpendicular to the plane and the conductivity tensors of the islands $\hat{\sigma}_1$ and the S2DG $\hat{\sigma}_2$. The task is solved for boundary condition that DC current $j_0$ is set on the infinity. The goal of the theory is to obtain conductivity tensor of the inhomogeneous system $\hat{\sigma}_{eff}$.
From the general physical principles the conductivity tensor must have the following form:
\begin{equation}
\hat{\sigma}_{eff}=\left(
\begin{matrix}
	\sigma^e_{xx} & -\sigma^e_{xy} \\
	\sigma^e_{xy} & \sigma^e_{xx}
\end{matrix} \right)
\end{equation}
Thus, there are only two independent variables $\sigma^e_{xx}$ and $\sigma^e_{xy}$. These values have to be expressed through $\sigma^n_{xx}$, $\sigma^n_{xy}$ of islands (n=1) and S2DG (n=2) and geometrical factor $p=\frac{\pi R^2}{a^2}$ that denotes the fraction of the system occupied by the islands.

\begin{figure}[t]
\centerline{\includegraphics[width=0.85\linewidth]{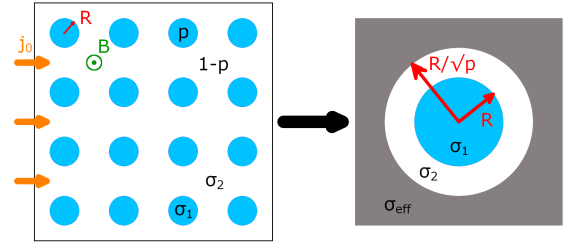}}
\caption{Schematic image of the transition of theoretical model according to mean field theory}
\label{Image}
\end{figure}

Such problems are very common for classical electrodynamics of continuous media and they can be solved within the widespread approach of mean field\cite{MFT}. In our case it claims that instead of periodical inhomogeneous system it is enough to consider the following system (see Fig.\ref{Image}): round island of radius $R$ with conductivity $\hat{\sigma}_1$ inside the ring with external radius $R_1=R/\sqrt{p}$ and conductivity $\hat{\sigma}_2$ that is surrounded by the effective medium with conductivity $\hat{\sigma}_{eff}$.

The dependence of electrical potential $\phi$ from coordinates is found using the continuity equation in stationary case div $\textbf{j}=0$, Ohm's law $\textbf{j}=\hat{\sigma}\textbf{E}$ and the definition of the potential $\textbf{E}=-\nabla\phi$. These conditions lead to the Laplace's equation $\Delta\phi=0$ where Laplace operator is taken in 2D. The solutions in different media are matched using the conditions of continuity of the electrical potential $\phi$ and radial component of the current $j_r$ on the borders. The solutions are chosen to satisfy the boundary conditions $j_x=j_0$ and $j_y=0$ at $x,y\rightarrow\infty$.

After all steps the electrical potential $\phi$ is expressed through the components of the conductivity tensor $\hat{\sigma}_{eff}$. Self-consistency conditions provide another one equation connecting these two values:

\begin{equation}
\hat{\sigma}_{eff}\cdot\overline{\textbf{E}}=\left( 
\begin{matrix}
j_0 \\ 0
\end{matrix} \right)
\end{equation}
where $\overline{\textbf{E}}$ is mean field:
\begin{equation}
\pi R_1^2 \overline{\textbf{E}} = \int^R_0 rdr\int_0^{2\pi}d\theta\textbf{E}(\textbf{r}) + \int^{R_1}_R rdr\int_0^{2\pi}d\theta\textbf{E}(\textbf{r})
\end{equation}
Finally, the following expressions for the components of $\hat{\sigma}_{eff}$ are obtained:
\begin{equation}
\label{Expres}
\begin{aligned}
\frac{\sigma^e_{xx}}{\sigma^2_{xx}}= 	& \left\{ \frac{(1-p)(1+\alpha^2)+\beta p(1-\alpha\gamma)}{(1-p)^2(1+\alpha^2)+\beta p[2(1-p)+\beta p]}+ \right. \\
																			& \left. +\frac{\gamma\sigma^2_{xy}}{2\sigma^2_{xx}}-\frac{1}{2} \right\} \frac{2}{1+\gamma^2} , \\
\sigma^e_{xy}=												& \;\gamma\sigma^e_{xx}, \\
\gamma = 															& \left\{ \frac{\sigma^2_{xy}}{2\sigma^2_{xx}}-\frac{\beta p\alpha}{(1-p)^2(1+\alpha^2)+\beta p[2(1-p)+\beta p]} \right\}\cdot \\
																			& \cdot \left\{ \frac{(1-p)(1+\alpha^2)+\beta p}{(1-p)^2(1+\alpha^2)+\beta p[2(1-p)+\beta p]}-\frac{1}{2} \right\}^{-1}
\end{aligned}
\end{equation}

where

\begin{equation}
\label{Express}
\begin{aligned}
\sigma^i_{xx}=\frac{\sigma_i}{1+(\mu_i B)^2}, \;&
\sigma^i_{xy}=\frac{\sigma_i \mu_i B}{1+(\mu_i B)^2}, \\
\alpha=\frac{\sigma^2_{xy}-\sigma^1_{xy}}{\sigma^2_{xx}+\sigma^1_{xx}}, \;&
\beta=\frac{2\sigma^2_{xx}}{\sigma^2_{xx}+\sigma^1_{xx}}
\end{aligned}
\end{equation}

Here $\sigma_i = n_i e\mu_i$ is Drude conductivity (i=1 corresponds to islands and i=2 to the S2DG), $p=\frac{\pi R^2}{a^2}\approx0.2$ (as $R$=2.5$\mu m$ and $a$=5$\mu m$). From these equations experimentally measurable value of the effective concentration $n_{eff}=\frac{B}{e\rho^e_{xy}}=B\frac{(\sigma^e_{xx})^2+(\sigma^e_{xy})^2}{e\sigma^e_{xy}}$ can be expressed. As a result $n_{eff} = n_{eff}(B,n_1,\mu_1,n_2,\mu_2)$, i.e. $n_{eff}$ depends on 5 parameters: magnetic field and concentration and mobility of the electrons in islands and S2DG. Fortunately, in our Si-MOSFET system both concentration $n$ and mobility $\mu$ of the electron gas are set by the voltage on the gate: $V_a$ for islands and $V_g$ for S2DG. This fact significantly simplifies analysis as effective concentration depends only on three variables: $n_{eff}(V_g,V_a,B) = n_{eff}(n_1(V_a),\mu_1(V_a),n_2(V_g),\mu_2(V_g),B)$.

\begin{figure}[t]
\centerline{\includegraphics[width=\linewidth]{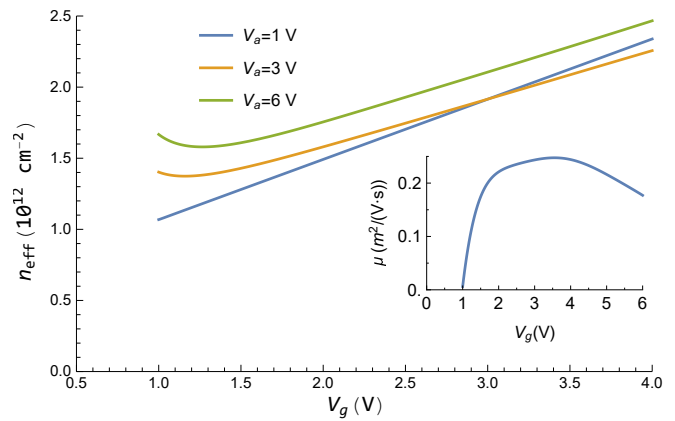}}
\caption{(Color online) Dependence of the effective concentration $n_{eff}$ obtained from the theory (eq.\ref{Expres}) from the voltage $V_g$ that parametrizes the matrix (S2DG) for three different voltages $V_a$ (1V, 3V and 6V) that parametrize the array of islands. The inset demonstrates the dependence of the mobility of electron gas from gate voltage which was used in theory.}
\label{TheoryNeffVg}
\end{figure}

The dependencies $n(V)$ and $\mu(V)$ were taken from the interpolation of the experimental data obtained on the conventional Hall bars from the same chip with investigated AA obtained in the same technological process: $n(V)$ is well approximated by a linear function, $\mu(V)$ - by a polynomial one. The experimental behavior of $n(V)$ and $\mu(V)$ differs strongly from one sample to another and, therefore, the coefficients in these functions shouldn't be considered as strict values defined by the samples. The dependence of $n_{eff}$ on $V_g$ for three different $V_a$ obtained from theoretical equations is shown in the Fig.\ref{TheoryNeffVg}. The inset of the Fig.\ref{TheoryNeffVg} demonstrates the dependence of the mobility of pristine electron gas on gate voltage $\mu(V)$. The dependence of concentration was taken $n(V)=0.645+0.4235\cdot V$ where $n$ and $V$ are measured in $10^{12} cm^{-2}$ and $V$, respectively. The magnetic field was taken to be 1T.

The given dependencies are very similar to the experimental dependence of $n_{eff}(V_g)$ shown in Fig.\ref{n(Vg)}. That is linear behavior for high values of $V_g$ and the bend of the graph for low $V_g$. The graph even demonstrates the low-$V_g$ upturn for high $V_a$. However here are also some distinctions of the theoretical model and the experimental graph. Firstly, in the Fig.\ref{TheoryNeffVg} graphs for low $V_a$ may intersect at some value of $V_g$ that never was observed in experiment. Secondly, on the theoretical dependence there is no bend down for low $V_a$. We attribute the emergence of both distinctions to the existence of shells. Theory doesn't take them into account at all, whereas they must crucially influence the system and the arise of drop for small $V_a$ and $V_g$ we attributed exactly to the enhanced role of shells.

To sum up, simple theoretical model given above satisfactorily describes the investigated system and reproduces the main features of the Hall effect behavior. For more strict description of the system and all the effects discovered experimentally the theoretical model should take into account the shells around the islands and quantum effects such as weak localization, SdHO. However taking these effects into account leads to the significant complication of the analytical result and the increase of the number of parameters. It can lead to impossibility of the analysis of such solutions.

\section{Discussion}
\subsection{Metal-insulator transition point}
Since Ioffe and Regel\cite{iofferegel} it is common knowledge that the boundary between metal and insulator corresponds to $k_Fl\sim 1$. In uniform 2D systems this criterion means that the conductivity is about $e^2/h\sim 1/26 $ kOhm . Below this value the wave functions at Fermi energy are localized and system is supposed to have insulating temperature dependence of the resistivity. Above this value the temperature dependence of the resistivity within non-interacting picture should be either weak insulating or metallic, in case of strong electron-electron interactions. The ultimate boundary between metal and insulator can be, of course, introduced only at $T=0$, when the coherence length is infinite. For macroscopic antidot array similar to ours, the low temperature limit can hardly be achieved, since it requires mK and sub-mK temperatures. S2DG is responsible for metal to insulator transition, while the geometrical factor (effective length-to-width ratio) in such system is enhanced. Therefore the threshold resistance per square in antidot array is elevated,  and 26 kOhm is not a dogma for macroscopically modulated system anymore. E.g. in our samples we observed vanishing temperature dependence of the resistivity for about 50-80 kOhm effective sample resistance.

\subsection{Phase diagram}
\begin{figure}[b]
\centerline{\includegraphics[width=\linewidth]{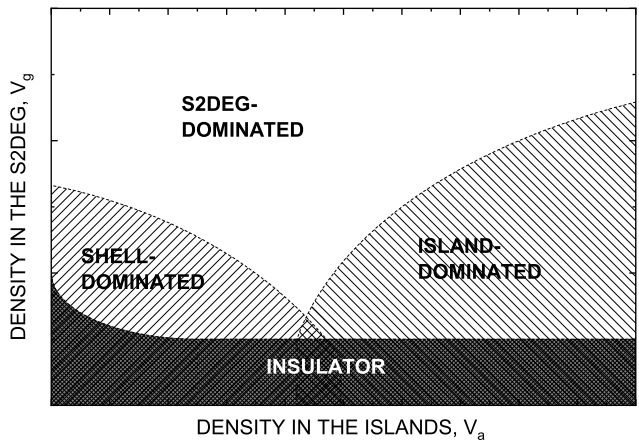}}
\caption{Schematic phase diagram of the system in space of S2DG (vertical) and islands (horizontal) electron density.}
\label{PhaseDiagram}
\end{figure}

Our results are summarized in the phase diagram of the system in $(V_g;V_a)$ plane in fig.\ref{PhaseDiagram}. For very low values of $V_g$ the system doesn't conduct, i.e. it is in insulating state. For low values of $V_g$ the value of $V_a$ is decisive. If $V_a$ is high enough, the system is in the island-dominated regime: current flows in the low-resistance islands and minimizes the path through the narrow bottlenecks between them. In this regime Hall density is elevated and Shubnikov-de Haas density is given by islands. Metallic temperature dependence of the conductivity is suppressed because total resistivity of the system is determined by bottlenecks between islands and S2DG.

For low values of $V_a$ the system is in the ``shell-dominated'' regime when current flows without preferences spreading out over the whole system. And for high values of $V_g$ again there is no big difference between low and high values of $V_a$ because islands are almost out of the game, the system is in the S2DG-dominated regime: current bypasses islands flowing through S2DG.

{\bf Role of periodicity}. Interestingly, the periodic structure (i.e. equivalence of all islands and inter-island necks) is important. In our case the period of the structure is 5 $\mu$m and there are only 80 periods across the 400 $\mu$m wide sample. If the system was more random, like e.g. \cite{minkov}, transport through it would be governed by percolation cluster and lateral cluster size could easily exceed 80 periods. In this case the properties of the system would be unreproducible and very large samples were needed for averaging, thus hindering the systematic studies.

\section{Conclusions}

To sum up, we experimentally examined transport properties of the macroscopically non-uniform and tunable Si-based 2D electron system and found the analytical solution of the simplified model system. Explored samples have two gates for controlling the densities in the islands and residual 2D gas separately. The conductive properties of this system turn out to depend on both gate voltages $V_g$ and $V_a$. The mean field theory gives a qualitative description of the experimental system including both gate voltages as parameters. In order to explain different behavior of the samples under different gate voltages we apply simple classical considerations about the current flow within 2DES. Finally, we suggest the phase diagram of the system, in coordinates electron density in the islands vs electron density in the 2D gas. In this phase diagram we identify various transport regimes from the analysis of the Hall effect and magnetoresistivity.


\section{Acknowledgements}
The authors are thankful to S.G. Tikhodeev, A.S. Ioselevich, L.E. Golub and V.Yu. Kachorovskii for discussions, and V.M. Pudalov for reading the manuscript.
The measurements were carried out using the equipment of the LPI Shared Facility Center. A.Yu. K. was supported by Basic research program of the HSE.

\end{document}